\documentclass[12pt]{iopart}
\usepackage{graphicx}
\usepackage{epsfig}
\usepackage{bm}
\usepackage{times}

\def\diff{\mathrm{d}}

\def\pra{Phys.\ Rev.\ A}
\def\prl{Phys.\ Rev.\ Lett.\ }

\def\beq{\begin{equation}}
\def\eeq{\end{equation}}

\def\reff#1{(\ref{#1})}

\def\subsc#1{{\mbox{\rm\scriptsize #1}}}

\def\Wcmcm{\mbox{\rm Wcm$^{-2}$}}

\def\TFWHM{T_\mathrm{FWHM}}

\def\N3d{N_\subsc{3D}}

\def\vekt#1{\bm{#1}}

\def\vektX{\vekt{X}}

\def\vektk{\vekt{k}}
\def\vektkhat{\hat{\vekt{k}}}

\def\vekte{\vekt{e}}
\def\vektE{\vekt{E}}

\def\vektA{\vekt{A}}

\def\Omegatilde{\tilde{\Omega}}

\def\twovec#1#2{\left( \!\! \begin{array}{c}\displaystyle #1 \\ #2 \end{array}\!\! \right)}
\def\twobytwomatr#1#2#3#4{\left( \!\! \begin{array}{cc} \displaystyle #1 & \displaystyle #2 \\\displaystyle #3 & \displaystyle #4 \end{array} \!\! \right)}

\def\halb{\frac{1}{2}}

\def\Edach{\hat{E}}

\def\Ehat{\Edach}

\def\energy{{\cal{E}}}

\def\abl#1#2{\frac{\diff #1}{\diff #2}}

\def\bra#1{\langle #1 \vert}
\def\ket#1{| #1 \rangle}

\def\imagi{\mathrm{i}}

\def\Deltatilde{\tilde{\Delta}}

\def\Up{U_\mathrm{p}}
\def\eulere{\mathrm{e}}

\def\OmegaR{\Omega_\mathrm{R}}
\def\Omegai{\Omega_\mathrm{i}}
\def\OmegaRtilde{\tilde{\Omega}_\mathrm{R}}

\begin{document}

\title{Nonperturbative resonant strong field ionization of atomic hydrogen}
\date{\today}
\author{M~G~Girju, K~Hristov, O~Kidun and D~Bauer}
\address{Max-Planck-Institut f\"ur Kernphysik, Postfach 103980, 69029 Heidelberg, Germany}
\date{\today}

\begin{abstract}
We investigate resonant strong field ionization of atomic hydrogen with respect to the 1s-2p-transition. By ``strong'' we understand that Rabi-periods are executed on a femtosecond time scale. Ionization and AC Stark shifts  modify the bound state dynamics severely, leading to nonperturbative signatures in the photoelectron spectra. We introduce an analytical model, capable of predicting qualitative features in the photoelectron spectra such as the positions of the Autler-Townes peaks for modest field strengths. {\em Ab initio} solutions of the time-dependent Schr\"odinger equation show a pronounced shift and broadening of the left Autler-Townes peak as the field strength is increased. The right peak remains rather narrow and shifts less. This result is analyzed and explained with the help of exact AC Stark shifts and ionization rates obtained from Floquet theory. Finally, it is demonstrated that in the case of finite pulses as short as $20$\,fs the Autler-Townes duplet can still be resolved. The fourth generation light sources under construction worldwide will provide bright, coherent radiation with photon energies ranging from a tenth of a meV up to tens of keV, hence covering the regime studied in the paper so that measurements of nonperturbative, relative AC Stark shifts should become feasible with these new light sources.     
\end{abstract}

\section{Introduction}
Resonant two-photon ionization of atomic hydrogen starting from the ground state  has been studied for more than thirty years as a prime example  for theoretical models in which two discrete states are coupled to the continuum (see, e.g.\  \cite{beers,crance77,acker78,knight79,austin79,geltman80,faisal81,mittle84} and references therein). The photoelectron spectra display duplets due to the AC Stark splitting of the bound-bound resonance (Autler-Townes \cite{autler} duplets). When plotted as a function of the detuning the energies of the two Autler-Townes peaks directly map the two field-dressed state energies into the continuum, thus making measurements of AC Stark shifts and avoided crossings possible \cite{feld92,shao95}. Besides this, it has been shown recently that the interference in Autler-Townes duplets can be controlled using two time-delayed intense femtosecond pulses \cite{wollen03}.  

If the coupling to the continuum is neglected, AC Stark shifts are ignored (or considered as input governing the effective detuning) and the rotating wave approximation is adopted, the well-known, analytically soluble Rabi-flopping dynamics are obtained. The influence of the latter on the emission spectra has been thoroughly studied for modest laser intensities (see Refs.~\cite{knight80,jentsch04} for reviews). More recently, harmonic generation from ionization-damped two-level systems \cite{pluci07}, radiation emitted by a resonantly driven H atom \cite{LaGattuta2} and the enhancement of intense-field high harmonic generation due to the coherent superposition of states \cite{milo06,zhang06}  have been studied theoretically.   

The unperturbed Rabi-flopping may be considered the zeroth order solution to be used for the calculation of spectra. Such approaches have been recently pursued to investigate ion impact ionization in the presence of a resonant laser \cite{voit06} and ionization of positronium in short, resonant UV laser pulses \cite{rod06}. This perturbative approach will work as long as ionization and AC Stark effect do not influence the bound state dynamics significantly. Our current work aims at studying  resonant strong field ionization beyond perturbation theory. On an {\em ab initio} basis and for short pulses this can be only achieved by a full numerical solution of the time-dependent Schr\"odinger equation (TDSE) \cite{LaGattuta1}.   

Our work is inspired by the  so-called fourth generation light sources now under construction worldwide (e.g.\  at DESY, Germany, in the UK and in the USA).  These new light sources  will provide coherent, short-wavelength radiation at intensities where the assumption of an unperturbed bound state flopping is invalid, opening up new possibilities to actually measure nonperturbative, relative AC Stark shifts of low-lying atomic or ionic states. First experiments on intense few-photon ionization of atoms have been already performed at the FLASH facility at DESY \cite{laar05,mosh07}.  

The paper is organized as follows. In Sec.~\ref{themodel} we introduce an analytical model, capable of describing the main qualitative features in the photoelectron spectra, which  are presented in Sec.~\ref{modelresults}. In Sec.~\ref{TDSE}  {\em ab initio} results from the solution of the TDSE are presented and compared with the model results. The nonperturbative features in the TDSE spectra are analyzed and explained with the help of Floquet theory in Sec.~\ref{floquet}. The case of strong field resonant ionization by femtosecond pulses is studied in Sec.~\ref{finitepulse}. Finally, we conclude in Sec.~\ref{concl}.

\section{Model} \label{themodel}
We consider a Hamiltonian of the form (atomic units are used unless specified otherwise)
\beq H(t)=H_0+W(t),\qquad H_0=T+V  \eeq
where $W(t)=z E(t)= z \Ehat(t) \cos\omega t$ accounts for the interaction with the linearly polarized, resonant (or almost resonant) laser field $\vektE(t)=E(t)\vekte_z$ in dipole approximation.
We approximate the solution of the TDSE \beq \imagi \ket{\dot{\Psi}(t)} = H(t)  \ket{\Psi(t)}   \label{tdse}\eeq
by using   the ansatz
\beq  \ket{\Psi(t)} = \eulere^{-\imagi\energy_a t} a(t) \ket{a} +  \eulere^{-\imagi\energy_b t} b(t) \ket{b} + \int\diff^3k\ c(\vektk,t) \ket{\vektk} . \label{ansatz}\eeq
Omission of the third term on the right hand side would lead us to the standard Rabi-theory (see any text book on quantum optics, e.g.\  \cite{zubscul}). 
The ansatz \reff{ansatz}, the neglect of continuum-continuum transitions $\sim \bra{\vektk'} V \ket{\vektk}$ (rescattering) and the assumption of mutual orthogonality of the states $\ket{a}$, $\ket{b}$, $\ket{\vektk}$ is equivalent to the use of the model Hamiltonian \beq {\cal{H}}(t)={\cal{H}}_0+{\cal{W}}(t) \eeq with
\beq {\cal{H}}_0  = \energy_a \ket{a}\bra{a} +  \energy_b \ket{b}\bra{b}  + \int\!\!\diff^3k\, \frac{k^2}{2} \ket{\vektk}\bra{\vektk} \eeq and
\beq {\cal{W}}(t) = {\cal{W}}_\mathrm{bb}(t) +  {\cal{W}}_\mathrm{bc}(t) +  {\cal{W}}_\mathrm{cc}(t) \eeq
where the subscripts ``bb'', ``bc'' and ``cc'' indicate bound-bound, bound-continuum and continuum-continuum coupling due to the laser field $E(t)$, respectively. The terms are explicitly given by
\begin{eqnarray*}
 {\cal{W}}_\mathrm{bb}(t) &=& E(t)  \bigl( \alpha \ket{b}\bra{a} + \mathrm{h.c.} \bigr) , \\
  {\cal{W}}_\mathrm{bc}(t) &=& E(t) \int\!\!\diff^3 k\, \bigl( d_a(\vektk) \ket{\vektk} \bra{a} + d_b(\vektk) \ket{\vektk} \bra{b} + \mathrm{h.c.} \bigr) , \\
 {\cal{W}}_\mathrm{cc}(t) &=& \imagi E(t)  \int\!\!\diff^3 k \diff^3k'\, \ket{\vektk} \partial_{k_z} \delta^{(3)}(\vektk-\vektk') \bra{\vektk'} 
\end{eqnarray*}
with the  transition matrix elements
\[ \alpha=\bra{b} z\ket{a},\qquad d_a(\vektk) = \bra{\vektk} z\ket{a},\qquad d_b(\vektk) = \bra{\vektk} z\ket{b}. \]
The model Hamiltonian ${\cal{H}}(t)$ is hermitian so that for, e.g., the initial conditions $a(0)=1$, $b(0)=c(\vektk,0)=0$ normalization $\vert a(t)\vert^2 + \vert b(t) \vert^2 + \int\diff^3k\, \vert c(\vektk,t)\vert^2=1 $ is ensured for all times.

The three coupled equations governing the time-evolution of the coefficients $a(t)$, $b(t)$ and $c(\vektk,t)$ read
\begin{eqnarray}
\imagi \dot{a}(t) &=& E(t) \left( \alpha^* b(t) \eulere^{-\imagi(\energy_b-\energy_a)t} + \eulere^{\imagi\energy_a t} \int\!\!\diff^3k\, c(\vektk,t) d_a^*(\vektk)  \right)\\
\imagi \dot{b}(t) &=& E(t) \left( \alpha a(t) \eulere^{\imagi(\energy_b-\energy_a)t} + \eulere^{\imagi\energy_b t} \int\!\!\diff^3k\, c(\vektk,t) d_b^*(\vektk)  \right)\\
\imagi \dot{c}(\vektk,t) &=& \frac{k^2}{2}  c(\vektk,t) + \kappa(a,b,\vektk,t) + \imagi E(t) \partial_{k_z}  c(\vektk,t)  \label{pde}
\end{eqnarray}
where
\[ \kappa(a,b,\vektk,t) = E(t) \left(  a(t) d_a(\vektk) \eulere^{-\imagi\energy_a t}  + b(t) d_b(\vektk) \eulere^{-\imagi\energy_b t}\right) .\]

The partial differential equation \reff{pde} can be formally integrated using the method of characteristics. The result reads
\beq c(\vektk,t) = -\imagi \int_{-\infty}^t\!\!\!\diff t'\ \kappa(a,b,\vektk+\vektA(t')-\vektA(t),t') \eulere^{-\imagi S_{\vektk}(t,t')} \label{cresult} \eeq
with $\vektA(t)=-\int^t \diff t'\ \vektE(t')$ the vector potential and $ S_{\vektk}(t,t')$ the action,
\beq  S_{\vektk}(t,t') = \halb\int_{t'}^t\diff t''\ [\vektA(t'')-\vektA(t)+\vektk]^2. \label{action}  \eeq

Introducing $\vektX=(a\eulere^{-\imagi\energy_a t},b\eulere^{-\imagi(\energy_b-\omega)t})^\top$, assuming $\alpha=\alpha^*$ being real and applying the rotating wave approximation (RWA) to the bound-bound dynamics lead to
\beq \imagi\dot{\vektX} = \twobytwomatr{\energy_a}{\frac{\OmegaR}{2}}{\frac{\OmegaR}{2}}{\energy_b-\omega} \vektX + E(t) \twovec{I_a(t)}{\eulere^{\imagi\omega t}I_b(t)} \label{system} \eeq
where
\begin{eqnarray}
\OmegaR &=& \Ehat \alpha, \\ 
I_a(t) &=&  \int\diff^3k\, d_a^*(\vektk) c(\vektk,t),\\
I_b(t) &=&  \int\diff^3k\, d_b^*(\vektk) c(\vektk,t) .  \end{eqnarray}
If the second term on the right hand side of Eq.~\reff{system} were known as an explicit function of time the differential equation could be solved analytically.
However, since $c(\vektk,t)$ depends via $\kappa(a,b,\vektk,t)$ on $a(t)$ and $b(t)$, Eq.~\reff{system} is an integro differential equation still too complicated to be solved analytically. Close to resonance and for modest ionization one may neglect in zeroth order the second term, obtaining the well-known unperturbed Rabi-flopping dynamics (see, e.g.\  \cite{zubscul})
\begin{eqnarray}
a_0(t) &=& \left[\cos(\Omega t/2) + \frac{\imagi\Delta}{\Omega} \sin(\Omega t/2)\right] \eulere^{-\imagi\Delta t/2} ,\label{rabia} \\
b_0(t) &=& \frac{\imagi\OmegaR}{\Omega}\sin(\Omega t/2) \eulere^{\imagi\Delta t/2}  \label{rabib}
\end{eqnarray} 
with
\beq \Omega = \sqrt{\OmegaR^2 + \Delta^2}, \quad \Delta=\energy_b-\energy_a-\omega, \label{rabifrequ} \eeq  
the Rabi-frequency and the detuning, respectively. We can then use \reff{rabia}, \reff{rabib} for $a(t)$ and $b(t)$ in $\kappa(a,b,\vektk,t)$ to calculate $c(\vektk,t)$ in first order according \reff{cresult} by simple numerical integration over time (for each $\vektk$ of interest). Subsequently, the photoelectron spectra can be calculated from the momentum distribution $P(\vektk)=\vert c(\vektk,t\to\infty)\vert^2$. The vector potential can be chosen such that $\vektA(t\to\infty)=\vekt{0}$. The probability for the emission of an electron with energy $\energy=k^2/2$ into the solid angle element $\diff\Omega_{\vektkhat}$ is then given by
\beq P_{\vektkhat}(\energy) = \frac{P(\vektk) \, \diff^3 k}{\diff\Omega_{\vektkhat}\,\diff\energy} = k P(\vektk) .\label{energyspec} \eeq

In the weak-field limit one may neglect the vector potentials in the argument of $\kappa$ and in the action \reff{action}. Equation \reff{cresult} then simplifies for $t\to\infty$ and $\Delta=0$ up to a constant phase to the first order perturbation theory (in $\omega$) result
\begin{eqnarray} c_0(\vektk) &=& \int_{-\infty}^\infty\!\!\diff t'\, E(t') \left( a_0(t') d_a(\vektk) \eulere^{-\imagi\energy_a t'} + b_0(t') d_b(\vektk) \eulere^{-\imagi\energy_b t'} \right) \eulere^{\imagi k^2 t'/2} \nonumber \\
&\simeq & \frac{\Ehat}{4} \int_{-\infty}^\infty\!\!\diff t'\, d_b(\vektk) \eulere^{\imagi t' (k^2/2 - \energy_b - \omega)} \left( \eulere^{\imagi\OmegaR t'/2} - \eulere^{-\imagi\OmegaR t'/2} \right) \label{firstorderpert}
\end{eqnarray} 
where the subscripts '$0$' indicate that the unperturbed Rabi-dynamics governed by Eqs.~\reff{rabia}, \reff{rabib} have been used.
From Eq.~\reff{firstorderpert} follows that a pair of peaks (a so-called Autler-Townes duplet) in the photoelectron spectra is expected at $\energy=\energy_b+\omega\pm {\OmegaR}/{2}$. Generalizing this result to higher photon orders we obtain 
\beq \energy=\energy_b+n \omega\pm\frac{\OmegaR}{2}, \quad n=1,2,\ldots . \label{peakpos} \eeq

\section{Model results}\label{modelresults}
For the case of atomic hydrogen, interacting with a linearly polarized laser field that is resonant (or almost resonant)  with the 1s $\leftrightarrow$ 2p$_0$-transition, i.e.\ $\energy_b-\energy_a=0.375=\omega+\Delta$, we calculated $ P_{\vektkhat}(\energy)$ according Eqs.~\reff{energyspec} using \reff{cresult} with  the  populations \reff{rabia}, \reff{rabib} and the dipole matrix elements for hydrogen-like ions of charge $Z$ (in our case $Z=1$),
\begin{eqnarray} d_a(\vektk) = d_\mathrm{1s}(\vektk) &=& -\imagi 2^{7/2} Z^{5/2} \frac{k_z}{\pi(k^2+Z^2)^3},\\ 
d_b(\vektk) = d_{\mathrm{2p}_0}(\vektk) &=& Z^{7/2} \frac{Z^2/4 + k^2 -6k_z^2}{\pi(k^2+Z^2/4)^4} . \end{eqnarray}
The Rabi-frequency at resonance is given by
\beq \OmegaR=\frac{256}{243} \frac{\Ehat}{\sqrt{2}} . \label{resonantrabi}\eeq

\begin{figure}[hbt]
\begin{center}
\includegraphics[width=0.7\textwidth]{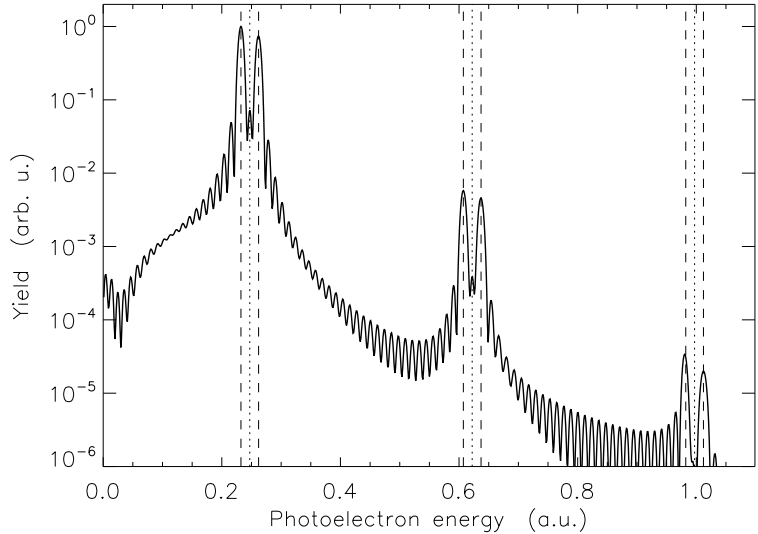}
\end{center}
\caption{Logarithmically scaled photoelectron spectra \reff{energyspec} in polarization direction $\vektkhat=\vekte_z$ for $\Ehat=0.04$ at resonance ($\omega=0.375$). For simplicity, a rectangular 32-cycle pulse was assumed. The atom undergoes $\simeq 2.5$ Rabi-cycles. The expected positions of the two Autler-Townes peaks \reff{peakpos}  are indicated by the dashed vertical lines, the center of them by the dotted vertical lines.  \label{resonantSFA} }
\end{figure}

Figure~\ref{resonantSFA} shows the spectrum at resonance ($\omega=0.375$) for  a  rectangular 32-cycle pulse of $\Ehat=0.04$. 
During such a pulse the atom undergoes $\simeq 2.5$ Rabi-cycles.  The positions of the Autler-Townes duplets, as expected from Eq.~\reff{peakpos}, are confirmed. 

\begin{figure}[hbt]
\begin{center}
\includegraphics[width=0.7\textwidth]{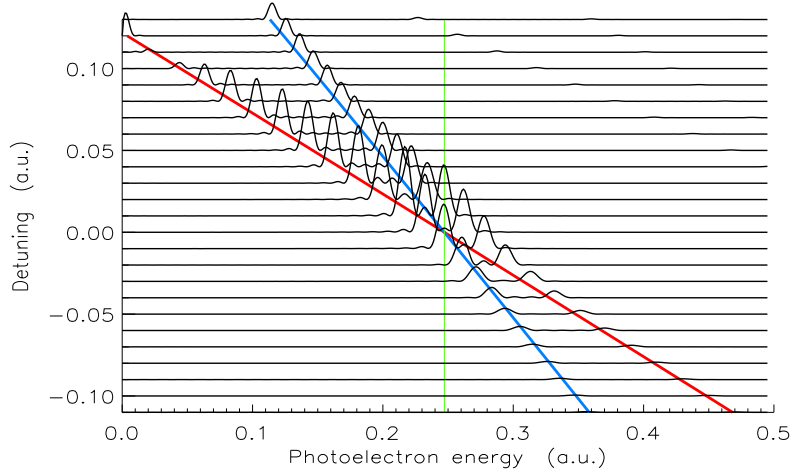}
\end{center}
\caption{ Spectra for various detunings $\Delta=0.375-\omega$ (other parameters as in Fig.~\ref{resonantSFA}).  The red, blue and green lines indicate $\energy_a+ 2\omega$,  $\energy_b+ \omega$ and $\energy_a+ 2\omega_\mathrm{res}$, respectively, with $\omega_\mathrm{res}=0.375$.   \label{detunedSFA} }
\end{figure}

It is instructive to trace the transition from the nonresonant to the resonant two-photon ionization. Figure~\ref{detunedSFA} shows the photoelectron spectra for various detunings $\Delta$ between the limits of double and single photon ionization, i.e.\ $\omega=0.25$ and $\omega=\vert\energy_a\vert=0.5$ or $\Delta=\pm 0.125$, respectively. The expected nonresonant lowest order peak positions follow straight lines and are given by $p_a(\omega) =  \energy_a + 2\omega$, $p_b(\omega) =  \energy_b + \omega$. These lines cross at resonance while the two actual photoelectron peaks do not. 
This is an experimentally observable manifestation of the well-known avoided crossings of field-dressed states \cite{feld92,shao95}. The two peaks appear to ``collide'' with (and repel) each other. Their separation at the closest approach (i.e.\  at resonance) equals the Rabi frequency $\OmegaR$ [cf.\ Eq.~\reff{peakpos}]. 

The model introduced in Sec.~\ref{themodel} relies on several simplifying assumptions. By taking the two-level dynamics \reff{rabia}, \reff{rabib} as the zeroth order solution to our problem of an infinite number of bound states plus a continuum coupled by a laser field and making the RWA, we neglect ionization losses, AC Stark shifts and anti-resonant terms.

It is well-known that ionization (or other losses) may be phenomenologically incorporated in the dynamics of a two-level system by introducing  complex energies and complex Rabi-frequencies (see, e.g.\  \cite{holt83} and references therein). In our model, this corresponds to the assumption that the second term on the right hand side of \reff{system} can be rewritten in the form 
\beq E(t) \twovec{I_a(t)}{\eulere^{\imagi\omega t}I_b(t)}  =  \twobytwomatr{\delta_a -\imagi \frac{\Gamma_a}{2}}{\imagi\frac{\Omegai}{2}}{\imagi\frac{\Omegai}{2}}{\delta_b-\imagi\frac{\Gamma_b}{2}} \vektX  \eeq
so that
\beq \imagi\dot{\vektX} = \twobytwomatr{\energy_a+\delta_a-\imagi\frac{\Gamma_a}{2}}{\frac{\OmegaR+\imagi\Omegai}{2}}{\frac{\OmegaR+\imagi\Omegai}{2}}{\energy_b-\omega+\delta_b-\imagi\frac{\Gamma_b}{2}} \vektX . \label{system2}\eeq
Here, $\delta_{a,b}$ are the AC Stark shifts and $\Gamma_{a,b}$ are the ionization rates.
Because of
\begin{eqnarray*} \abl{}{t} [ \vert a(t) \vert^2 + \vert b(t) \vert^2 ] &=& - \Gamma_a \vert a(t) \vert^2 - \Gamma_b \vert b(t) \vert^2  \nonumber \\
&& + 2 \Omegai \Re [a^*(t) b(t)\eulere^{-\imagi\Delta t}] \end{eqnarray*}
the bound state population then overall decreases due to ionization, provided $\Omegai$ is chosen sufficiently small.  In Ref.~\cite{holt83} the parameters $\delta_{a,b}$, $\Gamma_{a,b}$ and $\Omegai$  were determined from Floquet calculations for laser field strengths smaller than the ones of interest here.

\section{Results from the numerical {\em ab initio} solution of the TDSE} \label{TDSE}
\begin{figure}[hbt]
\begin{center}
\includegraphics[width=0.7\textwidth]{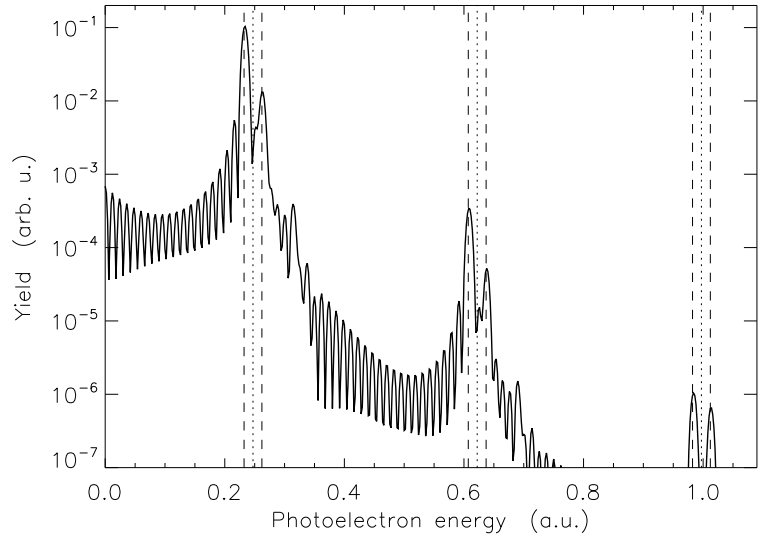}
\end{center}
\caption{TDSE electron spectra in polarization direction $\vektkhat=\vekte_z$ for $\Ehat=0.04$ and $\omega=0.375$, to be compared with the model result in Fig.~\ref{resonantSFA}. The laser pulse was 32 cycles long and of trapezoidal shape (up and down ramping over one cycle).  \label{resonantTDSE} }
\end{figure}

\begin{figure}[hbt]
\begin{center}
\includegraphics[width=0.7\textwidth]{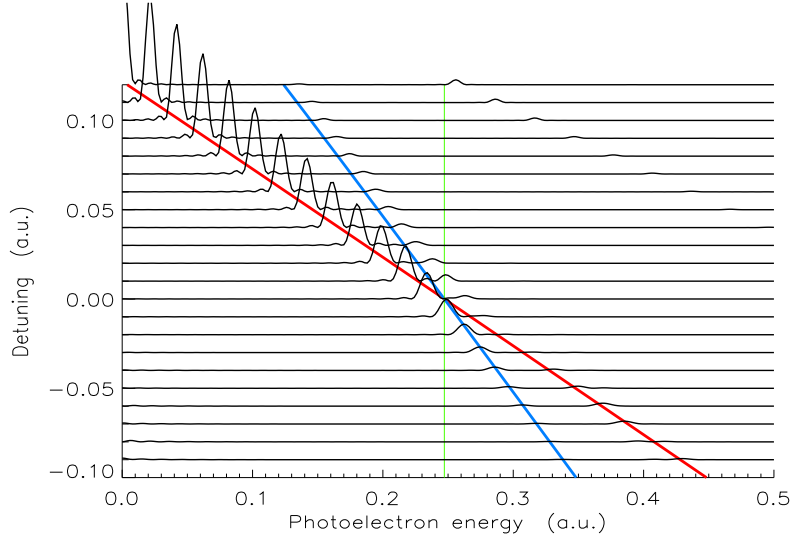}
\end{center}
\caption{ TDSE spectra for various detunings $\Delta=0.375-\omega$ (other parameters as in Fig.~\ref{resonantTDSE}), to be compared with the model result in Fig.~\ref{detunedSFA}.   \label{detunedTDSE} }
\end{figure}

The validity of the above model can be checked by comparing to {\em ab initio} solutions of the TDSE $\imagi\ket{\dot{\Psi}} = H(t) \ket{\Psi}$. We used the Qprop package \cite{qprop} to propagate the exact wave function and to calculate the photoelectron spectra. 
Figures \ref{resonantTDSE} and \ref{detunedTDSE} show the TDSE results corresponding to the results of Figs.~\ref{resonantSFA} and \ref{detunedSFA}, respectively. While the qualitative agreement is good as far as the peak positions are concerned, the relative strengths of the two peaks are different in the TDSE and the model calculations. Comparing Figs.~\ref{detunedSFA} and \ref{detunedTDSE}, it is clearly seen that the strength of the peak $p_b$ is overestimated in the model for positive detuning $\omega<\energy_b-\energy_a$. This is in agreement with observations made in Ref.~\cite{LaGattuta1} where also the left Autler-Townes peak was found to dominate the right peak.  Moreover, the TDSE results show that ionization from the 1s state prevails already for small positive detunings, justifying one of the assumptions of the so-called strong field approximation \cite{kfr} where all bound states different from the initial state are neglected.

\begin{figure}[hbt]
\begin{center}
\includegraphics[width=0.7\textwidth]{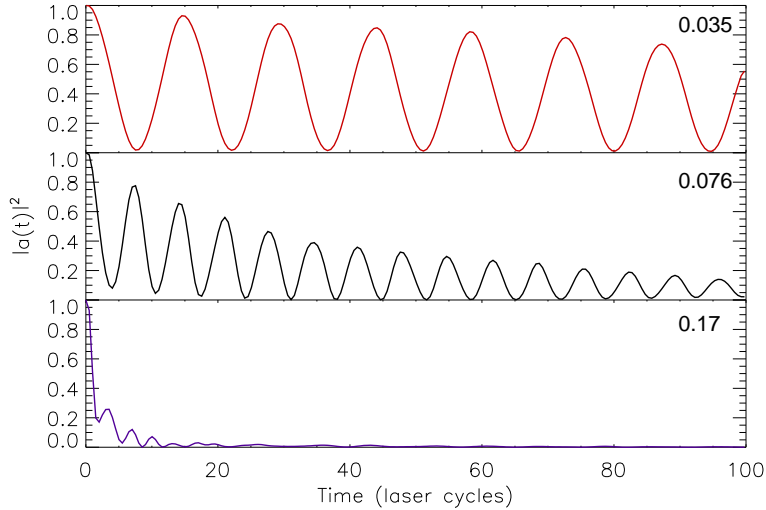}
\end{center}
\caption{\label{populations}  Ground state populations $\vert a(t)\vert^2$  vs time (in cycles of the driving laser  of resonant frequency $\omega=0.375$) for $\Ehat=0.035$, $0.076$ and $0.17$.}
\end{figure}

\begin{figure}[hbt]
\begin{center}
\includegraphics[width=0.75\textwidth]{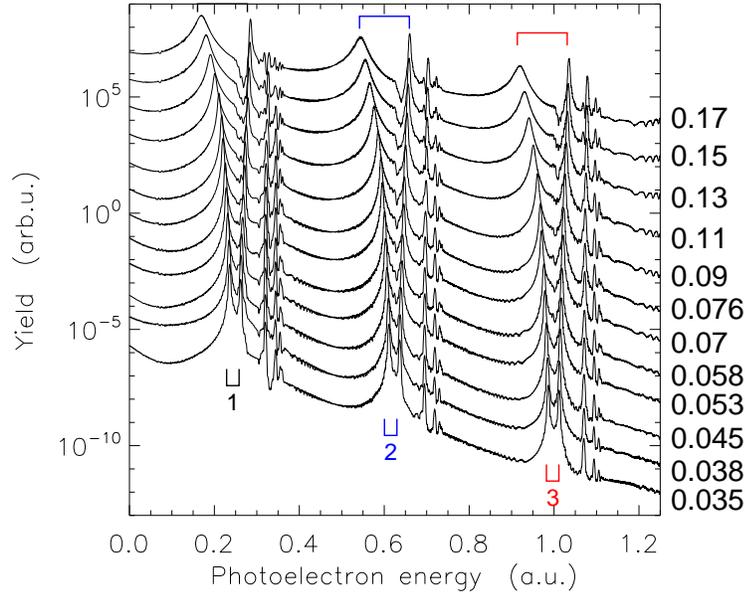}
\end{center}
\caption{\label{spectraforvariousEhat}   Angle-integrated photoelectron spectra for various laser field amplitudes $\Ehat$ (given to the right) for $\omega=0.375$. The spectra were vertically shifted for better visibility. The Autler-Townes duplets $i=1,2,3$ are indicated. }
\end{figure}

\begin{figure}[hbt]
\begin{flushright}
\includegraphics[width=0.85\textwidth]{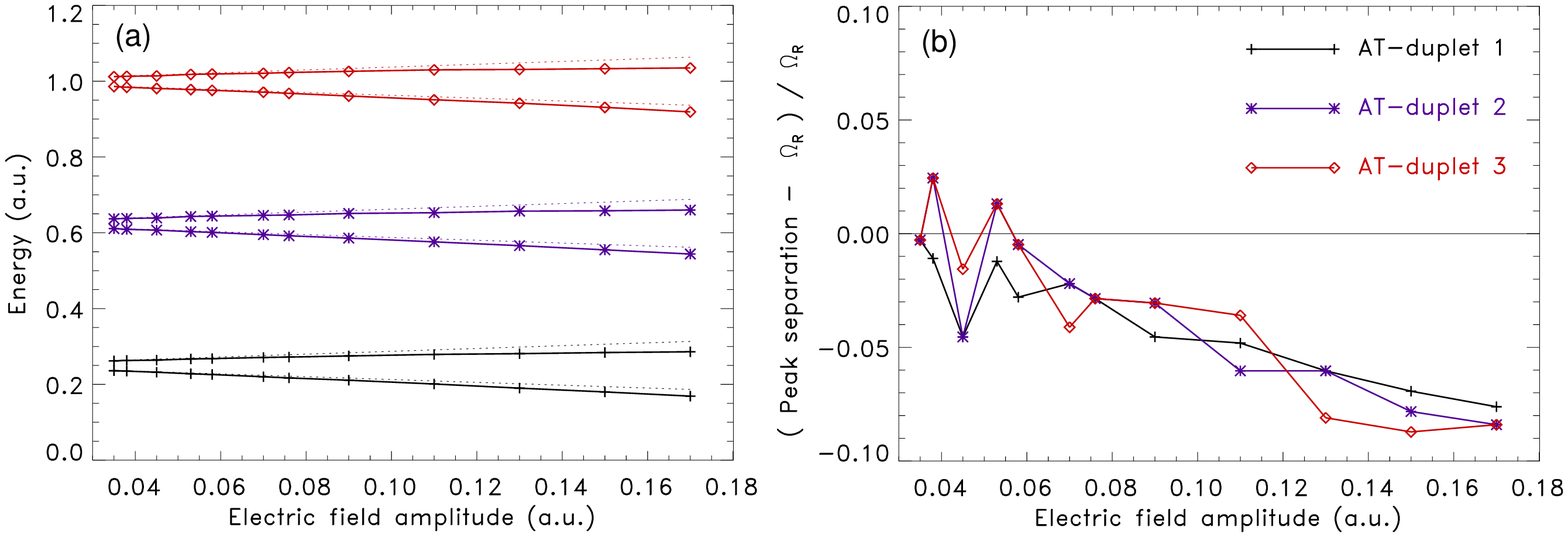}
\end{flushright}
\caption{\label{peak_diff}  (a) Peak positions of the first three Autler-Townes duplets vs electric field amplitude ($\omega=0.375$) as obtained from the TDSE solution (symbols) compared to the expected positions according $\energy_b+n\omega\pm\OmegaR/2$ with $n=1,2,3$ (dashed). (b) Relative peak separations  $(d_i-\OmegaR)/\OmegaR$ of the two Autler-Townes peaks in the $i$th Autler-Townes duplet for $i=1,2,3$.}
\end{figure}

\begin{figure}[hbt]
\begin{center}
\includegraphics[width=0.7\textwidth]{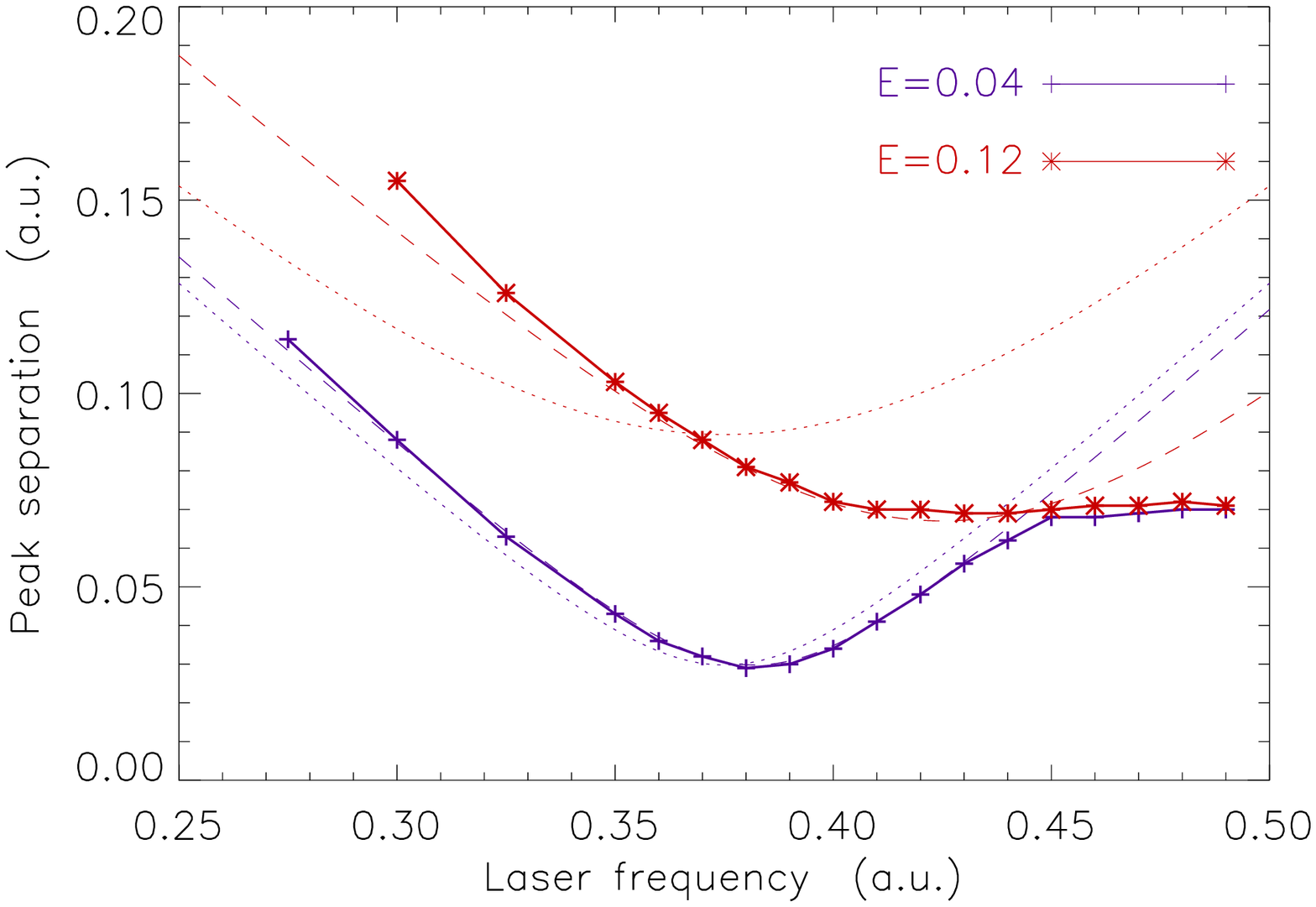}
\end{center}
\caption{\label{peakposdetuning}  Separation of the two peaks in the first Autler-Townes duplet (i.e.\  the ``true'' Rabi-frequency $\Omegatilde$) as a function of the laser frequency for $\Ehat=0.04$ (blue, +) and $\Ehat=0.12$ (red, *). The expected result according \reff{rabifrequII} (dotted) and the shifted parabolas  \reff{rabifrequIII} (dashed) are included. From the latter, the ``true'' detuning $\Deltatilde$ and resonant Rabi frequency $\OmegaRtilde$ can be determined. }
\end{figure}

Next, we investigate how the photoelectron spectra change when the atom is resonantly driven stronger and stronger, up to $\Ehat=0.17$ (corresponding to an intensity $\simeq 10^{15}$\,\Wcmcm). Here, by  ``resonant'' we understand a constant laser frequency of  $\omega=0.375$, which is the ``true'' resonance frequency only for negligible AC Stark effect and ionization, i.e.\  $\delta_{a,b}\simeq 0 \simeq \Gamma_{a,b} $. The  AC Stark effect is expected to push the system towards a positive detuning  $\Deltatilde = \energy_b + \delta_b - (\energy_a + \delta_a) - \omega > 0$ since $\delta_b > \delta_a$. Hence, the dynamic AC Stark effect leads to an increasing Rabi frequency. On the other hand, ionization losses damp the system, and damping is expected to shift the Rabi-flopping frequency to smaller values.

Both AC Stark shifts of {\em all} states (incl.\ the continuum) as well as ionization are included in the TDSE solution. 
In the following TDSE calculations a trapezoidal 100-cycle pulse was applied, ramped up and down over one cycle. For the lowest laser intensity shown ($\Ehat=0.035$) about $6.5$ Rabi-cycles occur during the pulse while at the highest intensity ($\Ehat=0.17$) ionization is already so violent that a Rabi-period cannot be identified unambiguously anymore. 
Figure~\ref{populations} shows the ground state populations $\vert a(t) \vert^2$  vs time for $\Ehat=0.035$, $0.076$ and $0.17$, respectively. While, as expected, the Rabi-frequency increases with increasing field strength, ionization leads to losses, i.e., the ground state population $\vert a(t) \vert^2$ does not return to unity. In the strongly over-damped case (bottom plot in Fig.~\ref{populations}) ionization clearly dominates the bound state dynamics. As long as $\Gamma_{a,b} \ll \OmegaR$ the oscillation frequency of the population transfer between the 1s and the 2p state is in good agreement  with the separation of the two peaks in an Autler-Townes duplet in the photoelectron spectra.
 We determined the separation $d_i$ of the two peaks within the $i$-th Autler-Townes duplet from the photoelectron spectra shown in Fig.~\ref{spectraforvariousEhat}. As expected, the separation increases with increasing laser field strength. Additional peaks originate from ionization involving higher-lying bound states. The left peak of each Autler-Townes duplet broadens as the field amplitude increases while the right Autler-Townes-peaks remain narrow. One may naively expect it to be the other way around because for increasing detuning the left (right) peak becomes the $\energy_a+2\omega$-peak  ($\energy_b+\omega$-peak) and ionization from the excited state $\ket{b}$ should be more probable, leading to a broadened {\em right} peak. Why the contrary is true will become clear in the next Section where we analyze the results in terms of Floquet theory.

 In Fig.~\ref{peak_diff} the (a) Autler-Townes peak positions and the (b) relative separations $(d_i-\OmegaR)/\OmegaR$ of the two peaks in the $i$-th Autler-Townes duplet for $i=1,2,3$ are plotted as a function of the driving laser field amplitude. With increasing driver strength the Autler-Townes duplet is shifted towards energies lower than those expected from the simple formula \reff{peakpos}. Figure~\ref{peak_diff}b shows that the peak separation $d_i$ becomes smaller than $\OmegaR$, indicating that the frequency down-shift due to ionization dominates the frequency up-shift due to the AC Stark shift detuning.

Figure~\ref{peakposdetuning} shows the separation $d_{i=1}$ of the two peaks in the first Autler-Townes duplet, i.e.\  the ``true'' Rabi-frequency $\Omegatilde$, as a function of the laser frequency for $\Ehat=0.04$ and $\Ehat=0.12$. In the modest intensity case  $\Ehat=0.04$ (corresponding to $5.6\times 10^{12}$\,\Wcmcm) the TDSE-result is close to the expected result according 
\beq \Omega(\omega) = \sqrt{\OmegaR^2+\Delta^2(\omega)} \label{rabifrequII} \eeq
(included dotted in Fig.~\ref{peakposdetuning}). A horizontal shift corresponding to the replacement
\beq \Delta \to \Deltatilde(\omega) = \energy_b + \delta_b - (\energy_a + \delta_a) - \omega = \Delta\energy-\omega+\delta_b-\delta_a \label{omegatilde} \eeq
with $\Delta\energy=0.375$ the unperturbed level spacing and $\delta_b-\delta_a=0.007$ yields the good agreement with the dashed curve $\Omegatilde(\omega) = \sqrt{\OmegaR^2+\Deltatilde^2(\omega)}$ in Fig.~\ref{peakposdetuning} for  $\Ehat=0.04$. For a pronounced negative detuning $\omega > 0.375 + \delta_b-\delta_a$ resonances with other (higher-lying) excited states come into play. In fact, in the TDSE spectra Autler-Townes duplets corresponding to the resonant coupling of the 1s ground state with the 3p ($\Delta\energy_{\mathrm{1s-3p}}=0.\overline{4}$) emerge when $\omega$ is closer to $0.\overline{4}$ than to $0.375 + \delta_b-\delta_a$.  

At higher laser intensities the Rabi frequency is shifted towards lower values owing to ionization (cf.\ Fig.~\ref{peak_diff}b). Hence, by fitting the TDSE results in Fig.~\ref{peakposdetuning} to
\beq \Omegatilde(\omega) = \sqrt{\OmegaRtilde^2+\Deltatilde^2(\omega)} \label{rabifrequIII} \eeq
with $\Deltatilde(\omega)$ defined in \reff{omegatilde} we can determine both  the relative AC Stark shift $\delta_b-\delta_a$ and the ``true'' resonant Rabi-frequency $\OmegaRtilde$.  In the strong field $\Ehat=0.12$-case ($5.1\times 10^{14}$\,\Wcmcm) a clear minimum (i.e.\  resonant Rabi-frequency $\OmegaRtilde$) and horizontal shift (i.e.\  $\Deltatilde-\Delta$) cannot be determined anymore.


\section{Floquet results} \label{floquet}
The TDSE results for long driving pulses may be best analyzed and understood in terms of the complex Floquet energies $\epsilon=\energy+\delta-\imagi\Gamma/2$ 
 which were introduced phenomenologically in Eq.~\reff{system2}. We used the STRFLO code \cite{STRFLO} to determine the exact Floquet energies. 

\begin{figure}[hbt]
\begin{center}
\includegraphics[width=0.6\textwidth]{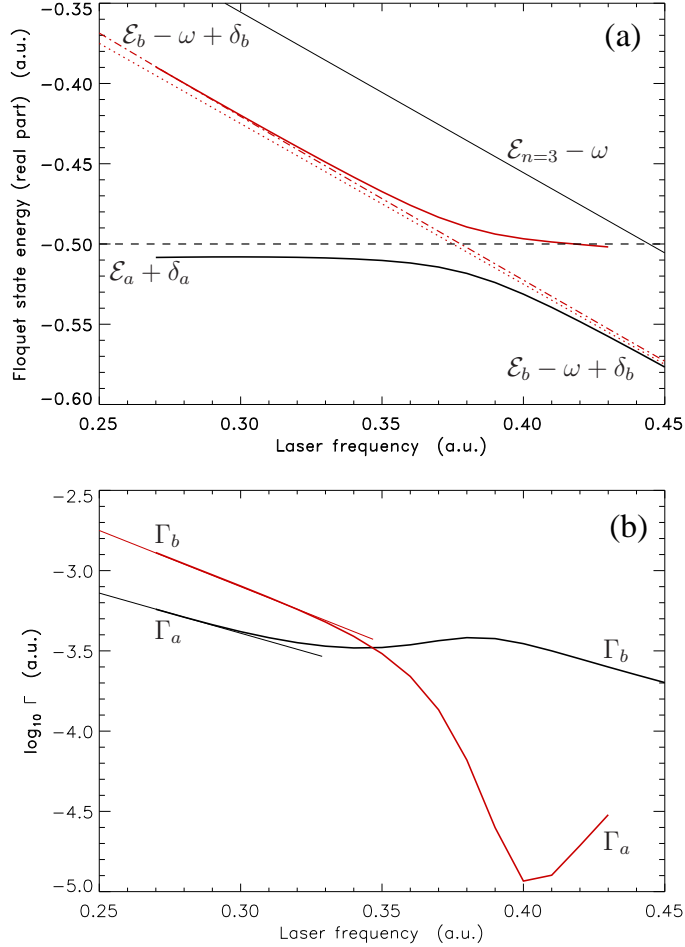}
\end{center}
\caption{ Real parts of the relevant Floquet energies as a function of the driver frequency $\omega$ for $\Ehat=0.04$ (a) and the corresponding ionization rates (b). \label{floquetenergies} }
\end{figure}

Figure~\ref{floquetenergies} illustrates how $\Re(\epsilon) = \energy+\delta$ and $\Gamma=-2\Im(\epsilon)$ (i.e.\  the ionization rates) behave as a function of the driver frequency for the modest field strength $\Ehat=0.04$. Well below the expected resonance at $\omega=0.375$ the energies follow $\energy_a+\delta_a$ and $\energy_b-\omega+\delta_b$. The AC Stark shift of the 1s state is negative while the 2p state is shifted upwards by $\simeq \Up$ [the dashed-dotted line is $\energy_b-\omega+\delta_b$ with $\delta_b=\Up(\omega)$]. Resonance occurs at the frequency where the two energies lie closest together. Since $\delta_b-\delta_a>0$ the resonance is shifted from the unperturbed value $\omega=0.375$ to the higher value $\simeq 0.38$. The line $\energy_{n=3}-\omega$ crosses $\energy_a=-0.5$ at the 1s$\leftrightarrow$3p resonance at $\omega=0.\overline{4}$.  In Fig.~\ref{floquetenergies}b the corresponding ionization rates do cross at $\omega \simeq 0.35$, i.e.\  already below resonance. For frequencies $0.25\leq\omega\leq 0.3$ the rates are well approximated by $\log_{10}\Gamma_a= -5\omega-1.89$ and $\log_{10}\Gamma_b= -7\omega-1$, respectively [straight lines in (b)]. For $\omega=0.375$ the ionization rate of the lower branch (drawn black  Fig.~\ref{floquetenergies}a) exceeds the ionization rate of the higher-lying red branch. This explains why the left peak  in an Autler-Townes duplet is broader than the right peak although the left peak is the one belonging to ionization from the 1s state  as one detunes the frequency towards smaller values. However, close to  resonance the dressed states are superpositions of the unperturbed states with sizable contributions from both $\ket{a}$ and $\ket{b}$.

\begin{figure}[hbt]
\begin{center}
\includegraphics[width=0.6\textwidth]{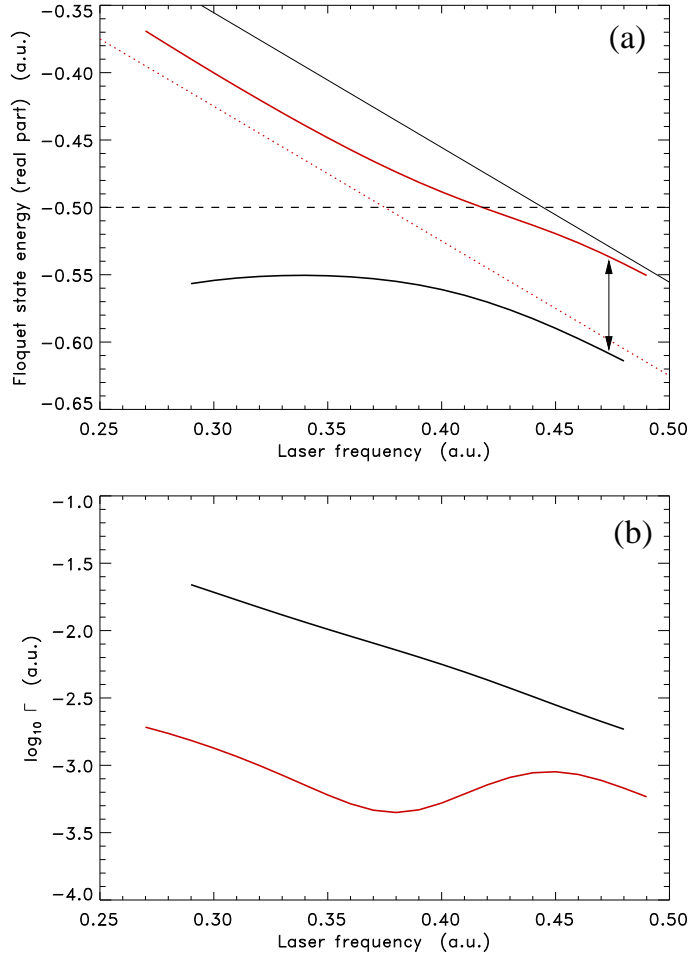}
\end{center}
\caption{ Same as in Fig.~\ref{floquetenergies} but for $\Ehat=0.12$. (a) With increasing frequency the branches approach $\energy_b-\omega$ (red, dotted) and $\energy_{n=3}-\omega$ (black, thin solid), respectively, with the mutual distance $\energy_{n=3}-\energy_b=0.069$ (indicated by the arrow) becoming constant. (b) The ionization rate of the lower branch is higher than that of the upper branch (adiabatic stabilization).    \label{floquetenergiesII} }
\end{figure}

Figure~\ref{floquetenergiesII} shows the Floquet energies and ionization rates for the stronger field  $\Ehat=0.12$.  The AC Stark shifts are very pronounced, shifting the resonance further towards higher frequencies where, however, the 1s$\leftrightarrow$3p resonance clearly affects the Floquet energy of the upper branch (red) whose AC Stark shift $\delta_b$ is less than $\Up$ for such high field strengths. With increasing frequency the branches approach $\energy_b-\omega$ (red, dotted) and $\energy_{n=3}-\omega$ (black, thin solid), respectively, with the mutual distance $\energy_{n=3}-\energy_b=0.069$ (indicated by the arrow) becoming constant. This explains why the peak separations obtained from the TDSE simulations shown in Fig.~\ref{peakposdetuning} both tend towards the value $\simeq 0.07$ with increasing frequency. The ionization rate of the lower branch is higher than that of the upper branch in the whole frequency range shown, not just close to resonance. This is a manifestation of so-called adiabatic stabilization \cite{gav02} which is known to set in for driving frequencies exceeding the ionization potential and for sufficiently high field strengths. The pronounced AC Stark down-shift of the lower branch and the high ionization rate associated with it are responsible for the respective shift towards lower energies and the broadening of the left Autler-Townes peak in the TDSE results of Fig.~\ref{spectraforvariousEhat}. An analysis of the Floquet wavefunctions corresponding to the energies shown in Fig.~\ref{floquetenergiesII} reveals that at such high field strengths more states than just $\ket{a}$ and $\ket{b}$ contribute in the frequency range of interest.

We have checked that the use of the AC Stark-shifted energies and ionization rates obtained from Floquet theory in Eq.~\reff{system2}  reproduces well the exact evolution of the populations $\vert a(t)\vert^2$ and   $\vert b(t)\vert^2$ from the TDSE. However, the quantitative agreement between the TDSE spectra and the model spectra calculated from \reff{cresult} using these exact populations is poor as far as the relative strengths of the Autler-Townes peaks and absolute figures for, e.g.,  the ionization probability are concerned. Giving up the RWA does hardly change the model results and thus does not improve the agreement, meaning that Bloch-Siegert shifts are of minor importance. The reason for the quantitative disagreement between model and TDSE results is most probably the plane wave ansatz in Eq.~\reff{ansatz} (instead of Coulomb continuum states).  It is known that the same approximation plagues the strong field approximation, and attempts to include Coulomb effects in the latter have been pursued (see, e.g.\  Refs.~\cite{smirn06,faisal06} and references therein).

\begin{figure}[hbt]
\begin{center}
\includegraphics[width=0.75\textwidth]{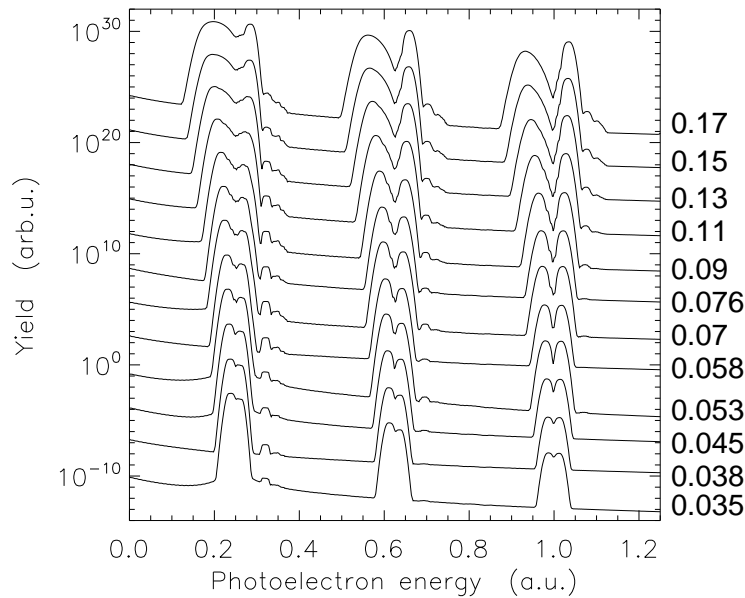}
\end{center}
\caption{Same as in Fig.~\ref{spectraforvariousEhat} but for a finite pulse $E(t)=\Ehat \exp[-4 (\ln 2) t^2/\TFWHM^2]\cos\omega t$ with $\TFWHM=20$\,fs.   \label{spectraforvariousEhatfinitepulse} }
\end{figure}

\section{Finite-pulse results} \label{finitepulse}
The new fourth generation light sources will deliver bright, coherent, short wavelength radiation as short {\em pulses}. It is clear that in cases where no Rabi-floppings can be completed within the pulse duration no clearly separated Autler-Townes peaks can be expected. According Eq.~\reff{resonantrabi} Rabi frequencies on the time-scale of a few femtoseconds are expected already at laser intensities $\simeq 3.5\times 10^{14}$\,\Wcmcm\ so that in a, say, $20$-fs pulse  many Rabi-cycles are executed and a clear separation of the two Autler-Townes peaks is anticipated. However, ionization and the AC Stark effect lead to a pronounced broadening of the left Autler-Townes peak at high intensities even for a flat-top pulse, as is visible in  Fig.~\ref{spectraforvariousEhat}. Hence the question arises whether in a real high-field, short-pulse experiment the Autler-Townes duplets can be resolved. Figure~\ref{spectraforvariousEhatfinitepulse} shows the TDSE results for exponential $20$-fs FWHM pulses (with respect to the electric field). Twenty femtoseconds is a realistic pulse duration for the new short wavelength radiation sources. The same peak field strengths as in Fig.~\ref{spectraforvariousEhat} were used so that flat-top and finite-pulse TDSE results can be directly compared. As to be expected, a broadening and smoothing due to the now time-dependent Rabi-frequency $\OmegaR=\OmegaR[\Ehat(t)]$ is visible. Nevertheless, the two Autler-Townes peaks are clearly separable, the contrast becoming better for higher order Autler-Townes duplets, presumably because of Coulomb effects having less influence on the more energetic photoelectrons.

\bigskip

\section{Conclusions} \label{concl}
We studied strong field ionization of atomic hydrogen close to the 1s-2p-resonance by means of an {\em ab initio} solution of the time-dependent Schr\"odinger equation. The photoelectron spectra directly reflect properties of the field dressed states, e.g.\  avoided crossings at resonances, different shifts of the two Autler-Townes peaks due to the highly nonperturbative AC Stark effect, broadening of peaks due to ionization and higher order Autler-Townes duplets. We found  ionization rates and AC Stark shifts hardly accessible to analytical theory for field strengths where the Rabi-dynamics occur on a femtosecond time scale. However, we were able to  explain the partly counter-intuitive results (such as the broadening of the left instead of the right Autler-Townes peak)  in terms of complex Floquet energies, giving the exact AC Stark shifts and ionization rates for infinite laser pulses. Finally, we showed that the Autler-Townes duplets remain resolvable in the spectra even when femtosecond pulses are applied. This is important in view of the fourth generation light sources under construction worldwide. We propose to use these novel sources of bright, coherent radiation to investigate strong field resonant ionization of low-lying atomic or ionic states experimentally in the nonperturbative regime.

\section*{Acknowledgment}
This work was supported by the Deutsche Forschungsgemeinschaft.


\end{document}